\documentclass[prl,twocolumn,showpacs,nofootinbib]{revtex4}
\usepackage{epsf}
\usepackage{amsmath}

\def\beq{\begin{equation}}
\def\eeq{\end{equation}}
\def\bea{\begin{eqnarray}}
\def\eea{\end{eqnarray}}
\def\beqa{\begin{equation}\begin{array}{l}}
\def\eeqa{\end{array}\end{equation}}
\def\eqlab#1{\label{eq:#1}}

\def\Eqref#1{Eq.~(\ref{eq:#1})}

\def\sla#1{#1 \hspace{-2mm} \slash}
\def\slap{p \hspace{-2mm} \slash}

\def\half{\mbox{\small{$\frac{1}{2}$}}}

\def\third{\mbox{\small{$\frac{1}{3}$}}}

\def\barr{\left(\begin{array}{c}}
\def\earr{\end{array}\right)}
\def\bmat{\left(\begin{array}{cc}}
\def\emat{\end{array}\right)}
\def\al{\alpha}
\def\be{\beta}
\def\ga{\gamma} 
\def\de{\delta} \def\De{\Delta}
\def\veps{\varepsilon}  \def\eps{\epsilon}

\def\la{\lambda} \def\La{{\Lambda}}

 \def\Si{{\it\Sigma}}

\def\pa{\partial}

\def\pa{\partial}

\def\nn{\nonumber}

\def\lag{{\mathcal L}}

\def\MM{{\mathcal M}}

\def\mathscr{\mathcal}

\def\3d{3-D}

\def\ol#1{\overline{#1}}

\def\ceft{$\chi$EFT}
\begin{document}
\preprint{WM-04-124}
\preprint{JLAB-THY-05-292}

\title{Magnetic moment of the $\Delta(1232)$-resonance in chiral
effective field theory
% observables and chiral extrapolation
}

\author{Vladimir Pascalutsa}
\email{vlad@jlab.org}
\author{Marc Vanderhaeghen}
\email{marcvdh@jlab.org}
\affiliation{Physics Department, The College of William \& Mary, Williamsburg, VA
23187, USA\\
Theory Group, Jefferson Lab, 12000 Jefferson Ave, Newport News, 
VA 23606, USA}

\date{\today}

\begin{abstract}
We perform a relativistic chiral effective field theory calculation of
the radiative pion photoproduction ($\gamma p \rightarrow \pi^0 p \gamma'$)
in the $\De$-resonance region,  to next-to-leading order in the 
``delta-expansion''. 
This work is aimed at a model-independent extraction of the $\Delta^+$  
magnetic moment from new precise measurements 
of this reaction. It also predicts the chiral behavior of
$\Delta$'s magnetic moment, which can be used to extrapolate
the recent lattice QCD results to the physical point.
\end{abstract}

%\pacs{13.60.Fz - Elastic and Compton scattering.
%14.20.Dh - Proton and neutrons.
%25.20.Dc - Photon absorption and scattering}%

\maketitle
\thispagestyle{empty}

The $\De(1232)$-isobar is the most distinguished and well-studied 
nucleon resonance. However, such a fundamental property as
its  magnetic dipole moment (MDM) has thusfar escaped a
precise  determination.
The problem is generic to any unstable particle whose lifetime
is too short for its MDM to be measurable in the usual way through 
spin precession experiments. A measurement of the MDM of
such an unstable particle can apparently be done only indirectly, in a 
three-step process, where the particle is first produced, then
emits a low-energy photon which plays the role of an external magnetic field, 
and finally decays. 
In this way the MDM of $\Delta^{++}$ is accessed in  
the reaction $\pi^+ p \to \pi^+ p \gamma$~\cite{Nef78,Bos91} while the
MDM of  $\Delta^+$ can be determined using the radiative pion photoproduction 
($\gamma p \to \pi^0 p \gamma^\prime$)~\cite{Drechsel:2000um}.

A first experiment devoted to the MDM of $\De^+$ 
was completed in 2002~\cite{Kotulla:2002cg}. The value extracted in this experiment,
$\mu_{\Delta^+} =  2.7 {{+1.0} \atop {-1.3}}
(\mathrm{stat.}) \pm 1.5 (\mathrm{syst.}) \pm 3 (\mathrm{theor.}) $ [nuclear magnetons],   
 is based on theoretical input from 
the phenomenological model~\cite{Drechsel:2001qu,Chiang:2004pw}
 of the $\gamma p \to \pi^0 p \gamma^\prime$  reaction. 
To improve upon the precision of this measurement, a dedicated series of experiments has recently 
been carried out by the Crystal Ball Collaboration at MAMI~\cite{CB}. These experiments achieve
about two orders of magnitude better statistics than the pioneering experiment\cite{Kotulla:2002cg}. 
The aim of the present work is to complement 
these high-precision measurements with an
accurate and model-independent analysis of the $\gamma p \to \pi^0 p \gamma^\prime$ reaction,
within the framework of chiral effective field theory (\ceft).

The \ceft\ of the strong interaction
is indispensable, at least at present, in relating the low-energy observables
(e.g., hadron masses, magnetic moments, scattering lengths)  
to {\it ab initio} QCD calculations on the lattice. On the other hand,
\ceft\ can and should be used in extracting various hadronic properties
from the experiment. In this Letter we will show how 
\ceft\ fulfills both of these roles in a gratifying fashion.
The one-loop calculation we present here is sufficient to
both complete the next-to-leading order calculation
of the $\gamma p \rightarrow \pi^0 p \gamma'$ reaction, in the
$\De$-resonance region, and perform a chiral
extrapolation of lattice QCD results for $\De$'s MDM~\cite{Lein91,Lee}. 

Our starting effective Lagrangian is that of the chiral perturbation theory ($\chi$PT) with pion and nucleon
fields~\cite{GSS89}. The $\De$ then is included explicitly in the so-called $\de$-expansion
scheme~\cite{PP03}. We organize the Lagrangian $\lag^{(i)}$, such that superscript $i$ stands for 
the power of electromagnetic 
coupling $e$ plus the number of derivatives
of pion and photon fields. Writing here only the terms involving the spin-3/2 isospin-3/2
field $\psi^\mu$
of the $\De$-isobar gives:\footnote{Here we introduce totally antisymmetric products
of $\ga$-matrices:
$\ga^{\mu\nu}=\half[\ga^\mu,\ga^\nu]$,
$\ga^{\mu\nu\al}=\half \{\ga^{\mu\nu},\ga^\al\}=
i\veps^{\mu\nu\al\be}\ga_\be\ga_5$. }
\begin{subequations}
\eqlab{lagran}
\bea
\lag^{(1)}_\De &=&  \ol\psi_\mu \left(i\ga^{\mu\nu\al}\,D_\al - 
M_\De\,\ga^{\mu\nu}\right) \psi_\nu \nn\\
&+& \!\frac{i h_A}{2 f_\pi M_\De}\left\{
\ol N\, T_a \,\ga^{\mu\nu\la}\, (\pa_\mu \psi_\nu)\, D_\la \pi^a 
+ \mbox{H.c.}\right\} \\
\lag^{(2)}_\De &=&  \frac{i e (\mu_\De-1)}{2M_\De}\, \ol \psi_\mu \psi_\nu\, F^{\mu\nu} \nn\\
&+& \frac{3 i e g_M}{2M (M+M_\Delta)}\left\{\ol N\, T_3
\,\pa_{\mu}\psi_\nu \, \tilde F^{\mu\nu}+ \mbox{H.c.}\right\} \nn\\
&-& \!\frac{e h_A}{2 f_\pi M_\De}\left\{
\ol N\, T_a\,\ga^{\mu\nu\la} A_\mu \psi_\nu\, \pa_\la \pi^a + \mbox{H.c.}\right\}\\
\lag^{(3)}_\De &=&  \frac{-3 e g_E}{2M (M+M_\Delta)}\left\{\ol N T_3
\ga_5 \pa_{\mu}\psi_\nu F^{\mu\nu}+ \mbox{H.c.}\right\}
\eea
\end{subequations}
where $M$ and $M_\De$ are, respectively, the nucleon and $\De$-isobar masses,
$N$ and $\pi^a\,\, (a=1,2,3)$ stand for the nucleon and pion fields, $D_\mu$ is the covariant 
derivative ensuring the electromagnetic gauge-invariance, $F^{\mu\nu}$ and $\tilde F^{\mu\nu}$
are the electromagnetic field strength and its dual,
$T_a$ are the isospin 1/2 to 3/2 transition matrices, coupling constants $f_\pi = 92.4$ MeV,
$g_M=2.94$, $g_E= -0.96$,
see~\cite{PP03,PP03b} for further details. The MDM $\mu_\De$ is defined here in units of
$[e/2M_\De]$. We omit the higher electromagnetic moments, because they do not contribute at the
orders that we consider.

Note that $\lag^{(1)}_\De$ contains the free Lagrangian, which is formulated in~\cite{RaS41}
such that the number of spin degrees of freedom of the relativistic spin-3/2 field is constrained
to the physical number: $2s+1=4$.  The $N$ to $\De$ transition couplings in \Eqref{lagran}
are consistent with these constraints~\cite{Pas98,PaT99}. 
The  $\ga \De\De$ coupling is more subtle since in this case
constraints do not hold for sufficiently strong electromagnetic fields, see, e.g., \cite{DPW00}. 
We do not deal with this problem here, thus assuming the electromagnetic field to be weak,
compared to the $\Delta$ mass scale.  

We now briefly describe the power counting in the $\de$-expansion scheme. 
The excitation energy of the $\De$-resonance, i.e., $\De\equiv M_\De-M_N\simeq 293$ MeV
is treated as a light scale,  so $\De\ll\La$, where $\La\sim 1$ GeV stands for 
the heavy scales of the theory. At the same time, $\De$ is counted differently from the other
light scale of the theory -- the pion mass, $m_\pi$. Namely,
$\De/\La$ counts as one power of the small parameter $\de$ while  $m_\pi/\La$
counts as two powers of $\de$.
Each graph can then be characterized 
by an overall $\delta$-counting index $n$,
which simply tells us that the graph is of size $\de^n$. 
Because the theory has two distinct light scales ($m_\pi$ and $\De$) 
the $\de$-counting index depends on whether the characteristic momentum $p$ is  
in the low-energy region ($p\sim m_\pi$) or in the resonance
region ($p\sim \De$). 
In the low-energy region the index of a graph with $L$ loops, $N_\pi$ pion propagators, $N_N$
nucleon propagators, $N_{\De}$ $\Delta$-isobar propagators, and $V_i$ vertices of
dimension $i$ is $n=2 n_{\chi{\mathrm PT}} - N_\De$, where
 $ n_{\chi{\mathrm PT}}=\sum_i i V_i + 4 L  - N_N - 2 N_\pi $
is the index in $\chi$PT with no $\De$'s \cite{GSS89}.

In the resonance region, one needs to distinguish the one-$\De$-reducible (O$\De$R) graphs,
because they contain $\De$ propagators which go as $1/(p-\De)$ and hence such
graphs are large and all need to be included. Their resummation amounts to 
dressing the $\De$ propagators so that they behave as $1/(p-\De-\Si)$. The self-energy 
$\Si$ begins at order $p^3$ and thus, for $p\sim\De$, 
the dressed O$\De$R propagator goes as $1/\de^3$. If the number of
such propagators in a graph is $N_{O\De R}$, the power-counting index of
this graph in the resonance region is given by 
$n=n_{\chi{\mathrm PT}} - N_\De - 2N_{O\De R}$.

\begin{figure}
\centerline{  \epsfxsize=8.5cm
  \epsffile{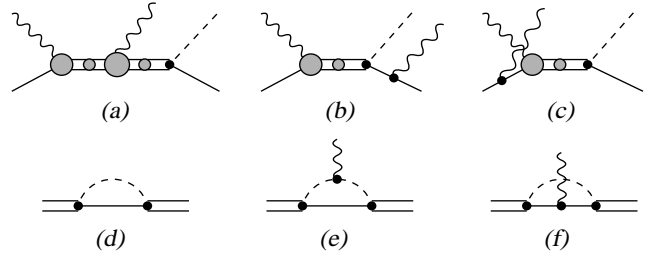} 
}
\caption{Diagrams for the $\gamma p \to \pi^0 p \gamma^\prime$ reaction 
at NLO in the $\delta$-expansion, considered in this work. Double lines represent
the $\De$ propagators.}
\label{diagrams}
\end{figure}

Consider now the amplitude for the $\gamma p \to \pi^0 p \gamma^\prime$ reaction. 
The optimal sensitivity to the MDM term is achieved when the incident photon energy
is in the vicinity of $\De$, while the outgoing photon energy is  of order of $m_\pi$.
 In this case the  $\gamma p \to \pi^0 p \gamma^\prime$ amplitude to next-to-leading
order (NLO) in the $\de$-expansion is given by the diagrams \ref{diagrams}(a), (b), and (c),
where the shaded blobs, in addition to vertices from \Eqref{lagran},
 contain the one-loop corrections shown in Fig.~\ref{diagrams}(d), (e), (f).
To present our calculation of the loops we introduce the  mass ratios:
$\mu=m_\pi/M_\De $, $r=M/M_\De$.  
The self-energy,  Fig.~\ref{diagrams}(d), has the following form
\begin{subequations}
\eqlab{selfen}
\beq
\Si^{\mu\nu}(\slap) =  A(p^2)\ga^{\mu\nu\al}\,p_\al + B(p^2) \ga^{\mu\nu},
\eeq
where the scalar functions, after dimensional regularization, take the form
\bea
A(p^2) &=& -\half C^2 \int_0^1 \! dx\, x\, {\cal M}^2\,  ( L-1 + \ln \MM^2), \\
B(p^2) &=& -\half C^2 r  \int_0^1 \! dx {\cal M}^2 \, ( L-1 + \ln \MM^2),
\eea
\end{subequations}
with $C=h_A M_\De/(8\pi f_\pi)$,  $L= - 2/(4-d) + \gamma_E + \ln (4 \pi M_\De/\La)$, 
$d \rightarrow 4$ the number of dimensions,
$\gamma_E =- \Gamma'(1) \simeq 0.5772 $, $\La$ the renormalization scale and
\beq
\MM^2(x) = x \mu^2 + (1-x) r^2 -x(1-x) (p^2/M_\De^2) - i\eps.
\eeq

After the on-mass-shell wave-function and mass renormalizations
the NLO $\De$-propagator is given by
\eqlab{prop}
\beq
S_{\mu\nu}(p) =\frac{-{\mathcal P}^{(3/2)}_{\mu\nu}(p)}{(\slap -M_\De)[1-i\mbox{Im}\,\Si'(M_\De)] 
- i\mbox{Im}\,\Si(M_\De)},
\eeq 
where we have introduced the spin-3/2 projection operator, ${\cal P}^{(3/2)}$, and
the following definitions: $\Si(M_\De)=M_\De A(M_\De^2) + B(M_\De^2)$, 
$\Si'(M_\De)=A(M_\De^2) + 2M_\De (\pa/\pa p^2) [M_\De A(p^2)+B(p^2)]_{p^2=M_\De^2}$.
These functions are complex when the $\De$-isobar is heavier than
the pion production threshold, $M_\De > M+m_\pi$. In this case $\MM^2 <0$ 
and hence the logarithm in \Eqref{selfen} gives rise to an imaginary part:
\begin{subequations}
\bea
{\rm Im}\, \Si(M_\De) &=& - (2\pi/3) M_\De C^2 (\al+r)\, \la^3 \,,\\
{\rm Im}\,\Si'(M_\De) &=& - 2\pi C^2\,
\la \left[ \al(1-\al)  (\al+r) \right. \nn\\
&-&\left. \third \la^2 (r+r^2-\mu^2)\right] ,
\eea
\end{subequations}
with $\al=\half(1+r^2-\mu^2)$,  $\la = \sqrt{\al^2 - r^2}$.
We note that the width of the resonance is given by 
$\Gamma_\De = - 2 \, {\rm Im}\, \Si(M_\De) $,
and find that the experimental value $\Gamma_\De\simeq 115$ MeV translates into $h_A\simeq 2.85$,
the value which we shall use in the numerical calculations.

The $\ga \De\De$ vertex, omitting the electric quadrupole and magnetic octupole terms,
is written in the form:
\bea
& & \bar u_\al (p') \, \Gamma^{\mu\al\be} (p',p) \, u_\be (p) \,\veps_\mu  \nn\\
& & =
e\, \bar u_\al (p')  \left[ \sla{\veps}\, F(q^2)  +\frac{(p'+p) \cdot \veps}{2M_\De}
\, G(q^2) \right]  u^\al (p) \,,
\eea
where $\veps^\mu$ is the photon polarization vector, $u^\al$ is the vector-spinor of the $\De$. 
In this notation, the MDM  is given by $\mu_\De = F(0)$.
The Ward-Takahashi identity, 
\beq
q_\mu  \Gamma^{\mu\al\be} (p',p) = 
e\left[(S^{-1})^{\al\be}(p')-(S^{-1})^{\al\be}(p)\right],
\eeq
demands that $F(0)+G(0)=1-\Si'(M_\De)$. This condition is verified explicitly in our calculation.
Since we have already given the expression for the self-energy, we present here only the 
expressions for $G$. The contributions of diagrams 1$(e)$ and 1$(f)$, at $q^2=0$, are:
\bea
&& G^{(e)}(0) =  -C^2 \int_0^1 \! dx\, x(1-2x)(x-r)\nn\\
&& \hskip2mm \times \left\{L +\ln[x \mu^2 + (1-x) r^2 -x(1-x)- i\eps]\right\} ,\\
&& G^{(f)}(0)  =  - 2 C^2 \int_0^1 \! dx\, x^2 (1-x -r) \nn\\
&& \hskip2mm \times \left\{ L +\ln[x r^2 + (1-x) \mu^2 -x(1-x)- i\eps]\right\}.
\eea
Their contribution to $\mu_\De$ for different isospin states is, e.g.,
$\mu^{(loop)}_{\De^{++}} =  F^{(e)} + F^{(f)}$, $\mu^{(loop)}_{\De^+} =  (1/3) F^{(e)} + (2/3) F^{(f)}$.

\begin{figure}[t,b,h]
%  \epsfxsize=5cm
%  \epsffile{xmuDe1.eps} 
\centerline{  \epsfxsize=6.5cm
  \epsffile{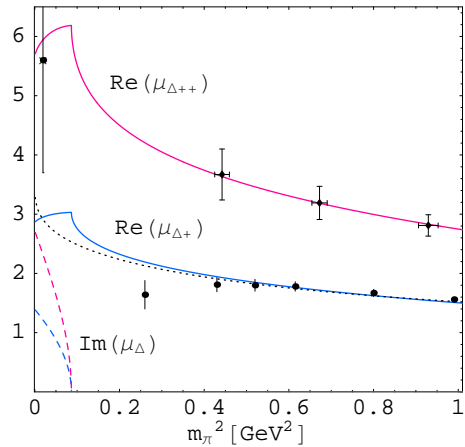}
}
\caption{Pion mass dependence of the real (solid curves) and imaginary
(dashed curves) parts of $\Delta^{++}$ and $\Delta^{+}$ MDMs [in nuclear magnetons].
Dotted curve is the result for the proton magnetic 
moment from Ref.~\cite{Pascalutsa:2004ga}. The experimental data point for $\De^{++}$
is from PDG analysis~\cite{PDG02}. Lattice data are from~\cite{Lein91} for $\De^{++}$ 
and  from~\cite{Lee} for $\De^+$.  }
\label{chiral}
\end{figure}

%[The $\pi N\De$
%coupling $h_A$ is fixed from the $\De$  width ($h_A\simeq 2.85$)
% while the two $\ga N\De$
%couplings are adjusted ($g_M=3$, $g_E=-1$) to reproduce]
As all lattice data for $\mu_\Delta$ at present and in the foreseeable 
future are for larger than the physical values of $m_\pi$, their comparison with experiment
requires the knowledge of the $m_\pi$-dependence for this quantity.
In contrast to the heavy-baryon result~\cite{Butler:1993ej}, our \ceft\ calculation
is manifestly relativistic, and, as argued earlier
for the nucleon~\cite{Pascalutsa:2004ga}, should be better suited for such extrapolations of lattice data.
Figure~\ref{chiral} shows the pion mass dependence of real and
imaginary  parts of the $\Delta^+$ and $\Delta^{++}$ MDMs, according to 
our one-loop calculation. Each of the two solid curves has a free 
parameter, the counterterm $\mu_\De$ from $\lag^{(2)}_\De$,
adjusted to agree with the lattice data at larger values of $m_\pi$.
As can be seen from 
Fig.~\ref{chiral}, the $\Delta$ MDM develops an 
imaginary part when $m_\pi<\Delta = M_\Delta - M$, 
whereas the real part has a pronounced cusp at $m_\pi = \Delta$. For $\mu_{\De^+}$
our curve is in disagreement with the trend of the recent lattice data,
which possibly is due to the ``quenching'' in the lattice calculations. 
The dotted line in Fig.~\ref{chiral} shows the result~\cite{Pascalutsa:2004ga}
for the magnetic moment for the proton. One sees that 
$\mu_{\Delta^+}$ and $\mu_p$, while having very distinct behavior
for small $m_\pi$, are approximately equal for larger values of $m_\pi$. 

\begin{figure}
\centerline{
  \epsfxsize=8cm
  \epsffile{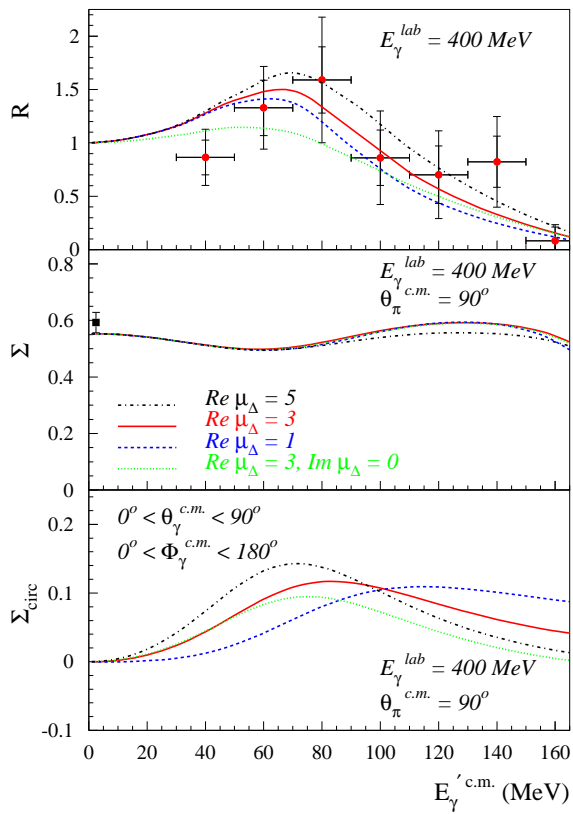}
}
\caption{The outgoing photon energy dependence of the 
$\gamma p \to \pi^0 p\ga'$ observables for different values of $\mu_{\Delta^+}$ 
(in units $e/2 M_\Delta$). 
Top panel: the ratio of $\gamma p \to \pi^0 p \gamma^\prime$ to 
$\gamma p \to \pi^0 p$ cross-sections Eq.~(\ref{eq:R1}). 
Data points are from~\cite{Kotulla:2002cg}. 
Middle panel: the linear-polarization photon asymmetry 
of the $\gamma p \to \pi^0 p \gamma^\prime$  
cross-sections differential 
w.r.t.\ the outgoing photon energy and pion c.m. angle. 
The data point at $E^\prime_\gamma = 0$ corresponds with the 
$\gamma p \to \pi^0 p$ photon asymmetry from~\cite{Beck:1999ge}.
Lower panel: the circular-polarization photon 
asymmetry (as defined in~\cite{Chiang:2004pw}), 
where the outgoing photon angles have been integrated over the 
indicated range. }
\label{fig:cross}
\end{figure}

We next discuss our results for the $\gamma p \to \pi^0 p \gamma^\prime$ 
observables. The NLO calculation of this process 
in the $\delta$-expansion corresponds with the diagrams of 
Fig.~\ref{diagrams}. 
As outlined above, this calculation 
completely fixes the imaginary  part of the 
$\gamma \Delta \Delta$ vertex. It leaves $\mu_\De$ as only free parameter,
which enters as a low energy constant in $\lag^{(2)}$. Thus the real part of $\mu_{\Delta^+}$ 
is to be extracted from the $\gamma p \to \pi^0 p \gamma^\prime$ observables, some of 
which are shown in Fig.~\ref{fig:cross} for an incoming photon 
energy $E_\gamma^{lab} = 400$~MeV as function of the emitted photon 
energy $E_\gamma^{\prime \, c.m.}$. In the soft-photon limit 
($E_\gamma^{\prime \, c.m.} \to 0$), the 
$\gamma p \to \pi^0 p \gamma^\prime$ 
reaction is completely determined from the bremsstrahlung process 
from the initial and final protons.  
The deviations of the $\gamma p \to \pi^0 p \gamma^\prime$ observables, 
away from the soft-photon limit, will then allow to study the 
sensitivity to $\mu_{\Delta^+}$. It is therefore very useful to 
introduce the ratio~\cite{Chiang:2004pw}:
\begin{eqnarray}
\label{eq:R1}
  R \,\equiv \, \frac{1}{\sigma_\pi} \cdot
  E^\prime_\gamma
  \frac{d\sigma}{dE^\prime_\gamma} ,
\end{eqnarray}
where $d\sigma / dE^\prime_\gamma$ is the 
$\gamma p \to \pi^0 p \gamma^\prime$ 
cross section integrated over the pion and photon angles, and 
$\sigma_\pi$ is the angular integrated cross section 
for the $\gamma p \to \pi^0 p$ process weighted with the bremsstrahlung 
factor, as detailed in~\cite{Chiang:2004pw}. 
This ratio $R$ has the property that in the soft-photon limit, the 
low energy theorem predicts that $R \to 1$. From 
Fig.~\ref{fig:cross} one then sees that the EFT calculation obeys this 
theorem. This is a consequence of gauge-invariance which is maintained exactly throughout
our calculation, also away from the soft-photon limit.
 
The EFT result for $R$ shows clear deviations 
from unity at higher outgoing photon energies, in good 
agreement with the first data for this process~\cite{Kotulla:2002cg}. 
The sensitivity of the EFT calculation to the $\mu_\Delta$ is a very 
promising setting for the dedicated second-generation experiment which 
has recently been completed by the Crystal Ball Coll.\ at MAMI 
\cite{CB}. It improves upon the statistics of the first experiment (Fig.~\ref{fig:cross})
by at least two orders of magnitude and will allow for a reliable extraction of 
$\mu_{\Delta^+}$ using the EFT calculation presented here. 
\newline
\indent
Besides the cross section, the $\gamma p \to \pi^0 p \gamma^\prime$ 
asymmetries for linearly and circularly polarized incident photons 
have also been measured in the recent dedicated experiment~\cite{CB}. 
They are also shown in Fig.~\ref{fig:cross}. 
The photon asymmetry for linearly polarized photons,  $\Sigma$, at $E^\prime_\gamma = 0$
exactly reduces  to the 
$\gamma p \to \pi^0 p$ asymmetry. It is seen from Fig.~\ref{fig:cross} that 
our calculation is in good agreement with the experimental value. 
At higher outgoing photon energies, the photon asymmetry
as predicted by the NLO EFT calculation 
remains nearly constant and is very weakly dependent on $\mu_\Delta$. 
It is an ideal observable for a consistency check of the EFT calculation 
and to test that the $\Delta$ diagrams of 
Fig.~\ref{diagrams} indeed dominate the process. Mechanisms involving $\pi$-photoproduction
Born terms followed by $\pi N$ rescattering have been considered in model 
calculations~\cite{Drechsel:2001qu,Chiang:2004pw}. In the $\de$-counting they
start contributing at next-next-to-leading order and therefore will provide the main source of
corrections to the present NLO results.
\newline
\indent
The asymmetry for circularly polarized photons, 
$\Sigma_{circ}$, (which is exactly zero for a 
two body process due to reflection symmetry w.r.t.\ the reaction plane) 
has been proposed~\cite{Chiang:2004pw} as a unique observable 
to enhance the sensitivity to $\mu_\Delta$. 
Indeed, in the soft-photon limit , where the 
$\gamma p \to \pi^0 p \gamma^\prime$ process reduces to a two-body process, 
$\Sigma_{circ}$ is exactly zero. 
Therefore, its value at higher outgoing photon energies 
is directly proportional to $\mu_\Delta$. 
One sees from Fig.~\ref{fig:cross} (lower panel) 
that our EFT calculation 
supports this observation, and shows sizeably different asymmetries 
for different values of $\mu_\Delta$. 
A combined fit of all three observables shown in Fig.~\ref{fig:cross} 
will therefore allow for a very stringent test 
of the EFT calculation, which can then be used 
to extract the $\Delta^+$ MDM. 

In conclusion, we have performed a manifestly gauge- and Lorentz-invariant chiral EFT
calculation for the $\gamma p \to \pi^0 p \gamma^\prime$ reaction
in the $\Delta(1232)$ resonance region. To next-to-leading order
in the $\delta$-expansion, the only free parameter entering the
calculation is the $\Delta^+$ magnetic dipole moment $\mu_{\Delta^+}$.
Due to the unstable nature
of the $\Delta$-isobar its magnetic moment acquires an imaginary part, an
effect which is computed in this work and will be taken into account in the extraction of  $\mu_{\Delta}$
from experiment.
Our present calculation is found to be in
good agreement with first experimental results for the
$\gamma p \to \pi^0 p \gamma^\prime$ cross sections, and will allow for
a model-independent extraction of $\mu_{\Delta^+}$ from a combined fit to
cross sections and photon asymmetries measured in new dedicated
experiments. At the same time, this chiral EFT calculation 
provides a crucial connection of present lattice QCD results for $\mu_\Delta$ at values
of $m_\pi > M_\Delta - M$ to the physical pion mass.

\begin{acknowledgments}
We thank Barry Holstein for interesting discussions.
This work is supported in part by DOE grant no.\
DE-FG02-04ER41302 and contract DE-AC05-84ER-40150 under
which SURA
operates the Jefferson Laboratory.  
\end{acknowledgments}

\end{document}